# Grainy Numbers

Gilles CHAMPENOIS

*Collège Saint-André, Saint-Maur, France*

*gilles_champenois@yahoo.fr*

ABSTRACT. Grainy numbers are defined as tuples of bits. They form a lattice where the meet and the join operations are an addition and a multiplication. They may be substituted for the real numbers in the definition of fuzzy sets. The aim is to propose an alternative negation for the complement that we'll call supplement.

> *"Die ganzen Zahlen hat der liebe Gott gemacht, alles andere ist Menschenwerk."*
>
> *Leopold Kronecker (1823-1891)*

## I. INTRODUCTION

In the now familiar Zadeh's theory [1], $k$-16 ($k$-$i$ here represents the number of years a student attend school before leaving) falls under *HIGH_EDUCATED* fuzzy set with a membership value of 1. In fuzzy set theory, the negation of a fuzzy set exists and is itself a set: for example $k$-16 (resp. $k$-8) falls under *NOT_HIGH_EDUCATED* fuzzy set with a membership coefficient of 0 (resp. 1). By contrast, *LOW_EDUCATED* is another fuzzy set which differs from *NOT_HIGH_EDUCATED* set because $k$-12 falls under the former with a coefficient of 0 and under the latter with a coefficient of more than $0,5$.

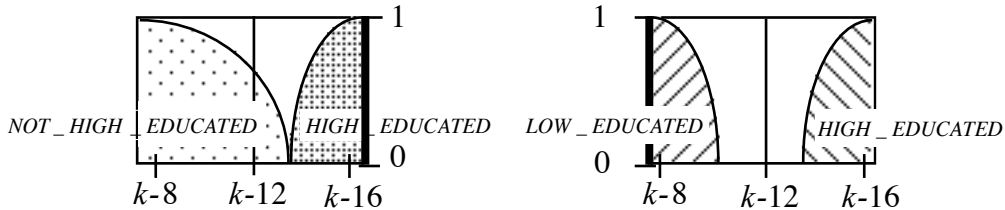

FIGURE 1. Two sorts of negation

We can see that two sorts of negation may be considered: the negation of complementation (*NOT_HIGH_EDUCATED* vs. *HIGH_EDUCATED*) which has been much studied in mathematical logic [2]; while the negation of contrariety (*LOW_* vs. *HIGH_EDUCATED*) has been put aside although ancient logic mentioned it. That kind of negation is precisely what we called supplementation negation. The intention here is to define grainy numbers and to substitute them for real numbers in the membership functions. So we could capture the notion of supplementation and put it on a mathematical track.

## II. DEFINITION

A grainy number is a tuple of bits with values among $\{-,1\}$.

DEFINITION. The symbols "$-$" (for example representing the empy set) and "1" (representing the power set of the empty set) will be used. Let $G^{\#}$ denote the set of tuples made of 1 and $-$. 0 denotes the empty tuple. Recursively we have,

<DEF $x \in G^{\#}$>
        <END $x = 0$>
        <NEXT $x = [1, x')$>
        <NEXT $x = [-, x')$>

EXAMPLE. $a = \left[1, \left[-, \left[1, 0\right)\right)\right) \in G^{\#}$ or simply $a = 1°3°$
            $b = \left[1, \left[-, \left[-, 0\right)\right)\right) \in G^{\#}$ or simply $b = 1°(3)$

| | | | | | |
|---|---|---|---|---|---|
| $[-, 0)$ | $-$ | $(1)$ | $\left[-, \left[-, \left[-, 0\right)\right)\right)$ | $- - -$ | $(3)$ |
| $[1, 0)$ | $1$ | $1°$ | $\left[1, \left[1, \left[1, 0\right)\right)\right)$ | $111$ | $1°2°3°$ |
| $\left[-, \left[1, 0\right)\right)$ | $-1$ | $2°$ | $\left[1, \left[-, \left[-, \left[-, 0\right)\right)\right)\right)$ | $1 - - - -$ | $1°(5)$ |
| $\left[1, \left[-, 0\right)\right)$ | $1-$ | $1°(2)$ | $\left[1, \left[-, \left[-, \left[1, 0\right)\right)\right)\right)$ | $1 - - - 1$ | $1°5°$ |
| $\left[1, \left[1, 0\right)\right)$ | $11$ | $1°2°$ | $\left[-, \left[-, \left[-, \left[-, \left[1, 0\right)\right)\right)\right)\right)$ | $- - - - 1$ | $5°$ |

$(1), (2), (3), \cdots$ are called flat grainy numbers.

*Addition*. We call addition the bitwise AND. If an operand is longer, the extra bits are cut. The addition of two grainy numbers $x$ and $y$ is a grainy number, written $x \oplus y$, and defined recursively as follows:

<DEF $z = x \oplus y = y \oplus x$>
        <END $x = 0$>                 $z = 0$
        <NEXT $x = [A, x')$ AND $y = [A, y')$>   $z = [A, x' \oplus y')$
        <NEXT $x = [1, x')$ AND $y = [-, y')$>   $z = [-, x' \oplus y')$

EXAMPLES.

$$3° \oplus 2° = (2)$$
$$1° \oplus 1° = 1°$$
$$1° \oplus (1) = (1)$$
$$1°3°4° \oplus 1°2°(4) = 1°(4)$$

The addition is associative and idempotent.

PROOF. Suppose that $x = \lceil A, x' \rceil$ and $x' \oplus x' = x'$. Then $x \oplus x = \lceil A, x' \oplus x' \rceil = \lceil A, x' \rceil = x$

*Ordering*. An order relation is defined on $G^{\#}$ as follows: $x \geq y$ iff $x \oplus y = x$.

EXAMPLES.
$$0 \geq 1° \geq 1°2° \geq 1°2°3° \geq 1°2°3°(5)$$

*Multiplication*. We call multiplication the biwise OR. If an operand is longer, the extra bits are concatenated. The multiplication of two grainy numbers $x$ and $y$ is a grainy number, written $x \otimes y$, and defined as follows.

$\langle \text{DEF } z = x \otimes y = y \otimes x \rangle$

|  |  |
|---|---|
| $\langle \text{END } x = 0 \rangle$ | $z = y$ |
| $\langle \text{NEXT } x = \lceil A, x' \rceil \text{ AND } y = \lceil A, y' \rceil \rangle$ | $z = \lceil A, x' \otimes y' \rceil$ |
| $\langle \text{NEXT } x = \lceil 1, x' \rceil \text{ AND } y = \lceil -, y' \rceil \rangle$ | $z = \lceil 1, x' \otimes y' \rceil$ |

EXAMPLES.
$$2° \otimes 3° = 2°3°$$
$$1° \otimes 1° = 1°$$
$$1° \otimes (1) = 1°$$
$$1°3°4° \otimes 1°2°(4) = 1°2°3°4°$$

The multiplication is associative and idempotent. Absorption laws hold: given two grainy numbers $x, y \in G^{\#}$, $y \oplus (x \otimes y) = y$ and $y \otimes (x \oplus y) = y$.

PROOF. We prove that $y \oplus (x \otimes y) = y$
  Suppose that $x = \lceil A, x' \rceil$ and $y = \lceil A, y' \rceil$
  $y \oplus (x \otimes y) = y \oplus \lceil A, x' \otimes y' \rceil = \lceil A, y' \oplus (x' \otimes y') \rceil = \lceil A, y' \rceil = y$
  Suppose that $x = \lceil -, x' \rceil$ and $y = \lceil 1, y' \rceil$
  $y \oplus (x \otimes y) = y \oplus \lceil 1, x' \otimes y' \rceil = \lceil 1, y' \oplus (x' \otimes y') \rceil = \lceil 1, y' \rceil = y$
  Suppose that $x = \lceil 1, x' \rceil$ and $y = \lceil -, y' \rceil$
  $y \oplus (x \otimes y) = y \oplus \lceil 1, x' \otimes y' \rceil = \lceil -, y' \oplus (x' \otimes y') \rceil = \lceil -, y' \rceil = y$

The multiplication distributes over the addition and vice versa.

$\left(G^{\#},\oplus,\otimes\right)$ is a distributive lattice.

## III. SUPPLEMENTATION

The supplementation is a binary operation that changes the grainy number $x$ in its bits specified by $k \in G^{\#}$. The supplementation of $x$ along $k$ is defined recursively as follows (We use $\overline{1}=0$ and $\overline{0}=1$),

$<\text{DEF } y = \overline{x}^{k}>$

| | |
|---|---|
| $<\text{END } x=0>$ | $y=0$ |
| $<\text{END } k=0>$ | $y=x$ |
| $<\text{NEXT } x=\left[X,x'\right] \text{ AND } k=\left[-,k'\right]>$ | $y=\left[X,\overline{x}'^{k'}\right)$ |
| $<\text{NEXT } x=\left[X,x'\right] \text{ AND } k=\left[1,k'\right)>$ | $y=\left[\overline{X},\overline{x}'^{k'}\right)$ |

EXAMPLE. $\overline{(3)}^{2^{\circ}} = 2^{\circ}(3)$

$\overline{1^{\circ}2^{\circ}3^{\circ}}^{2^{\circ}} = 1^{\circ}3^{\circ}$

Note that:

- the supplementation is involutive,
- $\overline{x}^{k} = x$ whenever $k$ is a flat number (for example, $\overline{1^{\circ}2^{\circ}3^{\circ}}^{(2)} = 1^{\circ}2^{\circ}3^{\circ}$),
- $\overline{x}^{x}$ is a flat number (for example, $\overline{1^{\circ}3^{\circ}}^{1^{\circ}3^{\circ}} = (3)$) called the module of $x$ and denoted by $|x|$,
- $\overline{|x|}^{x} = x$ for any $x \in G^{\#}$.

## IV. FUZZY SETS FROM GRAINY NUMBERS

In the historical definition of fuzzy sets, we intend to replace the unit interval by the grainy lattice. $\langle A,* \rangle \in G^{\#}$ will denote the membership function of the set $A$.

*Full set*. It is the set whose membership function equals $\langle A,* \rangle = 0$.

*Containment*. A set $A$ contains a set $B$ iff $\langle B,* \rangle \leq \langle A,* \rangle$.

*Reunion*. The reunion of two sets $A$ and $B$ is a set $A \cup B$, whose membership $\langle A \cup B,* \rangle$ equals $\langle A,* \rangle \oplus \langle B,* \rangle$.

*Intersection*. The intersection of two sets $A$ and $B$ is a set $A \cap B$ whose membership function $\langle A \cap B,* \rangle$ equals $\langle A,* \rangle \otimes \langle B,* \rangle$.

*Supplementation*. The supplement of a set $A$ along $k \in G^{\#}$ is the set (written $\overline{A}^{k}$) defined by $\left\langle \overline{A}^{k}, * \right\rangle = \overline{\left\langle A, * \right\rangle}^{k}$.

## V. EXAMPLE

EXAMPLE N°1. Let's consider the two sets *HIGH_EDUCATED* and *LOW_EDUCATED* defined by their respective membership functions:

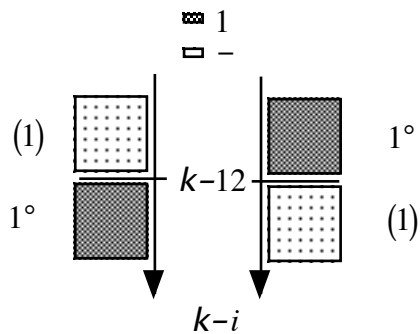

FIGURE 2. The grainy membership functions of *HIGH_EDUCATED* (left) and *LOW_EDUCATED* (right)

We see that *LOW – EDUCATED* is the first order supplement of *HIGH – EDUCATED*,

$$\overline{HIGH\_EDUCATED}^{1°} = LOW\_EDUCATED$$

EXAMPLE N°2. We now consider the sets *HIGH_EDUCATED* and *LOW_EDUCATED* extended to the sets *VERY_HIGH_EDUCATED* and *VERY_LOW_EDUCATED*:

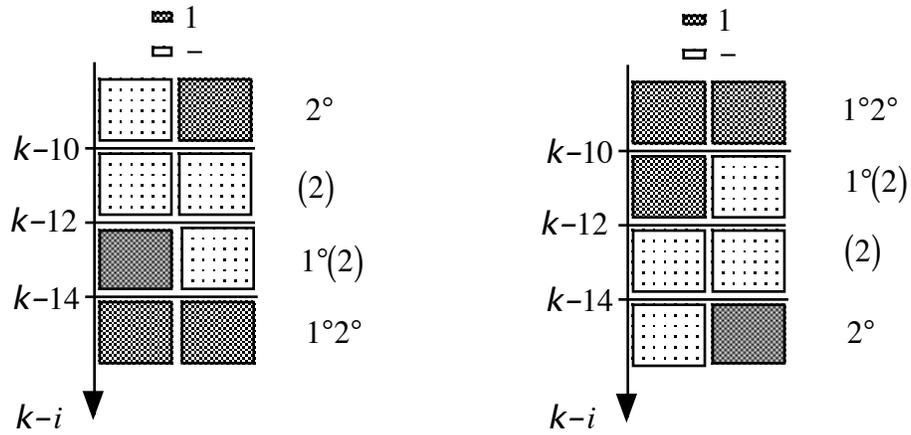

FIGURE 3. The grainy membership functions of *HIGH _ EDUCATED* (left) and *LOW _ EDUCATED* (right)

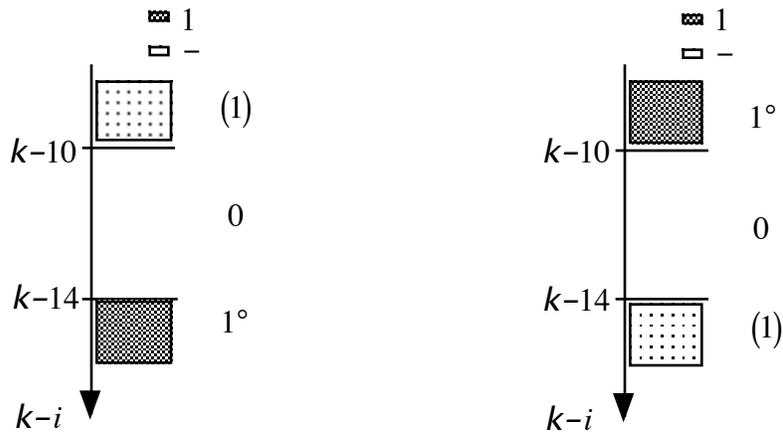

FIGURE 4. The grainy membership functions of *VERY _ HIGH _ EDUCATED* (left) and *VERY _ LOW _ EDUCATED* (right)

We see that *LOW – EDUCATED* is still the first order supplement of *HIGH – EDUCATED*,

$$\overline{HIGH\_EDUCATED}^{1°} = LOW\_EDUCATED$$

and *VERY _ LOW – EDUCATED* is also the first order supplement of *VERY _ HIGH – EDUCATED*,

$$\overline{VERY\_HIGH\_EDUCATED}^{1°} = VERY\_LOW\_EDUCATED$$